\documentclass[10pt, a4paper]{article}
\pdfoutput=1 
\usepackage{authblk}

\usepackage{cite}
\usepackage{url}
\usepackage{amsmath,amssymb,amsfonts}
\usepackage[]{algorithm}
\usepackage[]{algorithmicx}
\usepackage{graphicx}
\usepackage{textcomp}
\usepackage{xcolor}
\usepackage{gensymb}
\usepackage{units}
\usepackage{todonotes}
\usepackage{etoolbox}

\usepackage[bf]{caption}
\usepackage{subcaption}
\usepackage{enumitem}
\usepackage{threeparttable}

\usepackage[a4paper, total={184mm,239mm}]{geometry}
\def\BibTeX{{\rm B\kern-.05em{\sc i\kern-.025em b}\kern-.08em
    T\kern-.1667em\lower.7ex\hbox{E}\kern-.125emX}}

\newcommand{\blue}[1]{{\color{black}#1}}

\DeclareRobustCommand*{\authorrefmark}[1]{\raisebox{0pt}[0pt][0pt]{\textsuperscript{\footnotesize\ensuremath{\ifcase#1\or *\or \dagger\or \ddagger\fi}}}}

\providecommand{\keywords}[1]
{
  \small	
  \textbf{\textit{Keywords---}} #1
}

\let\svmaketitle\maketitle
\def\maketitle{\svmaketitle\thispagestyle{empty}}
    
\begin{document} 

\title{{\vspace{-2cm}\small Accepted for publication at the 24th Design Automation and Test in Europe Conference (DATE) 2021. \textcopyright 2021 IEEE.}\vspace{-0.8\baselineskip}
\rule{\textwidth}{0.4pt}\vspace{0.8cm}
\bf Reliability-Aware Quantization for Anti-Aging NPUs}
\date{}

\author[]{
Sami Salamin\authorrefmark{1},
Georgios Zervakis\authorrefmark{1}, 
Ourania Spantidi\authorrefmark{2},\\
Iraklis Anagnostopoulos\authorrefmark{2},
J{\"o}rg Henkel\authorrefmark{1},
and Hussam Amrouch\authorrefmark{3}
}
\affil[]{
\small
\textit{\authorrefmark{1}Chair for Embedded Systems (CES), Karlsruhe Institute of Technology, Karlsruhe, Germany}\\
\textit{\authorrefmark{2}Department of Electrical and Computer Engineering, Southern Illinois University, Carbondale, U.S.A.}\\
\textit{\authorrefmark{3}Chair of Semiconductor Test and Reliability (STAR), University of Stuttgart, Stuttgart, Germany}\\
\textit{\authorrefmark{1}\{sami.salamin, georgios.zervakis, henkel\}@kit.edu, \authorrefmark{2}\{ourania.spantidi, iraklis.anagno\}@siu.edu, \authorrefmark{3}amrouch@iti.uni-stuttgart.de}
}

\maketitle

\begin{abstract} 
Transistor aging is one of the major concerns that challenges designers in advanced technologies.
It profoundly degrades the reliability of circuits during its lifetime as it slows down transistors resulting in errors due to timing violations unless large guardbands are included, which leads to considerable performance losses.
When it comes to Neural Processing Units (NPUs), where increasing the inference speed is the primary goal, such performance losses cannot be tolerated.
In this work, we are the first to propose a reliability-aware quantization to eliminate aging effects in NPUs while completely removing guardbands. 
Our technique delivers a graceful inference accuracy degradation over time while compensating for the aging-induced delay increase of the NPU.
Our evaluation, over ten state-of-the-art neural network architectures trained on the ImageNet dataset, demonstrates that for an entire lifetime of 10 years, the average accuracy loss is merely 3\%.
In the meantime, our technique achieves 23\% higher performance due to the elimination of the aging guardband.

\end{abstract}

 
\section{Introduction}\label{sec:intro}
Late advancements in Neural Networks (NNs) research boosted the accuracy of several machine learning applications, at the cost of an immense increase in computational demands~\cite{jouppi2017datacenter}.
However, to achieve that and to bring the inference speed of Deep NNs (DNNs) to an acceptable level, custom ASIC Neural Processing Units (NPUs) are becoming ubiquitous in general purpose and embedded computing~\cite{jouppi2017datacenter,cass2019taking,song20197}.
An NPU consists of thousands of multiply-accumulate (MAC) units~\cite{our_esweek}, which provide massive parallelism of the performed computations that DNNs demand.
Google TPU integrates 64K MACs~\cite{jouppi2017datacenter}, while even the embedded-oriented Samsung~\cite{song20197} and Google~\cite{cass2019taking} NPUs employ 1K and 4K MACs, respectively. 
In NPUs, a large number of MAC units are tightly packed together within a small footprint.
Their inherent nature to simultaneously perform tens of tera operations per second, makes NPUs subject to elevated on-chip power densities that rapidly result in excessive on-chip temperatures during operation~\cite{our_esweek}.

\textbf{Circuit Aging:} The very high utilization of MAC circuits within NPUs~\cite{jouppi2017datacenter,our_esweek} exposes the underlying transistors to continuous stress with very little time for relaxation and recovery.
As a result, transistors age faster.
In addition, the presence of excessive temperatures, as mentioned earlier, exacerbates further the problem as the majority of mechanisms behind transistor aging exponentially depend on the operating temperature~\cite{bti_temperature,BTI_Overview_Mahapatra}.
Generated defects, due to transistor aging phenomena such as Bias Temperature Instability (BTI) and Hot-Carrier Injection (HCI), manifest themselves as a degradation in the main electrical characteristics of transistors.
In practice, the threshold voltage ($V_{th}$) increases~\cite{BTI_Overview_Mahapatra} leading, in turn, to a reduction of the drain current of a transistor in the ON state ($I_{on}$).
This considerably increases the propagation delay of the transistor and thus, of the logic cells (Eq.~\ref{eq:ion}-\ref{eq:inv_delay}~\cite{hu_book}). 
Hence, circuits exhibit timing errors because the operating frequency becomes unsustainable over time.
To overcome that and keep aging effects at bay for the entire projected lifetime (e.g., 10 years), a timing guardband ($t_{GB}$) must be included on top of the critical path delay (Eq.~\ref{eq:guardband}) at \textit{design time}. 
This leads directly to large losses in performance from the beginning until the end of lifetime even through aging-induced delay degradations do not yet exist (or are very small) at the early phases of the chip's lifetime.
Hence, the cost of guardbanding is paid from the very beginning even though it is not yet needed (Eq.~\ref{eq:perf_loss}). 
Several approaches have been proposed~\cite{aging_no_gb,tran_aging} w.r.t. guardband narrowing and aimed at minimizing the associated performance losses.
However, they reduce the aging impact by inducing area/power overhead, through transistor over-sizing~\cite{aging_no_gb}.

\begin{align}
&I_{on} \propto  V_{dd} - ( V_{th} + \Delta V_{th}) \label{eq:ion}  \\
&t_{CP}(fresh) =  \sum \limits_{m_i \in CP}^{} D_{m_i} \text{; }  
D_{m_i}  \approx \frac{C V_{dd}}{4}\left ( \frac{1}{I_{onN}} + \frac{1}{I_{onP}}\right )  \label{eq:inv_delay}  \\
& t_{freq}(x) <  t_{CP}(fresh) + t_{GB}(y),\ x \leqslant y   \Rightarrow \text{\textit{timing errors} \textbf{!}} \label{eq:guardband} \\
& t_{freq}(x) >  t_{CP}(fresh) + t_{GB}(y),\ x < y  \Rightarrow \text{\textit{Perf. loss} \textbf{!}} \label{eq:perf_loss} 
\end{align}

\noindent
where $x$ is chip's age, $x{=}0$ as fresh chip. $y$ is projected lifetime. $m_i$ refers to transistors that form the circuit's critical path ($CP$). 
$D_{m_i}$ is simplified propagation delay of logic gate~\cite{hu_book},
and $C$ represents the load capacitance connected to the gate.

\textbf{Aging-Aware Approximation:} Recently, approximate computing has been employed to address aging in error-tolerant applications~\cite{amrouch_approximate_aging, amrouch_approximate_aging2}.
Approximate computing exploits the inherent error resilience of several applications, to trade-off computational accuracy with other metrics, e.g., delay~\cite{amrouch_approximate_aging, amrouch_approximate_aging2, zervakis2020design, TasoulasTCASI2020, mrazek2019alwann}.
Aging-aware works in approximate computing introduce directed approximations to improve a circuit's performance and mitigate the aging effects.
However, they examined very simple topologies, e.g., RCA adders and array multipliers~\cite{amrouch_approximate_aging, amrouch_approximate_aging2}.
 
\textit{In this work, we suppress aging effects in NPUs by applying, for the first time, adaptive approximation through input compression in which reliability-aware quantization is used}.
With a marginal inference accuracy loss, we demonstrate that aging guardbands can be removed for the entire projected lifetime.

\noindent
\textbf{Our novel contributions within this paper are as follows:}\\
(1) This is the first work that employs quantization as a novel mechanism to eliminate aging effects in NPUs.\\
(2) We present, for the first time, a graceful-approximation technique that suppresses, over time, aging effects in NPUs. Our technique enables designers to remove aging guardbands and hence eliminates the associated performance loss.\\
(3) We demonstrate that for an average accuracy loss of merely 3\%, our technique eliminates the performance loss due to aging guardband, i.e., $23$\%, while timing errors caused by transistor aging are suppressed for the entire lifetime of 10 years.

\section{Related work}
\label{sec:related}
In~\cite{mrazek2019alwann,zervakis2020design,TasoulasTCASI2020} approximate multipliers are employed and run-time reconfigurable approximate NN inference accelerators are implemented.
However,~\cite{mrazek2019alwann,zervakis2020design,TasoulasTCASI2020} target power and not delay optimization.
In~\cite{amrouch_approximate_aging2}, fixed approximation through precision scaling is applied in order to narrow or remove aging guardbands.
Nevertheless, fixed approximation leads to constant quality degradation that cannot be adapted over time.
In~\cite{amrouch_approximate_aging}, adaptive input cutting and masking techniques are proposed to mitigate aging and achieve a graceful accuracy degradation of a DCT/IDCT accelerator.
However,~\cite{amrouch_approximate_aging} was only applied to the very slow ripple-carry adder and array multiplier.
Moreover, \cite{amrouch_approximate_aging} requires control circuitry to set the run-time approximation.
\cite{amrouch_approximate_aging,amrouch_approximate_aging2} reduce the computational precision of the accelerator itself.
Hence, considering that errors due to approximate hardware are input-dependent, by just omitting some bits from the computations~\cite{amrouch_approximate_aging,amrouch_approximate_aging2}, the quality loss for some inputs might be unacceptable~\cite{zervakis2020design}.

\begin{figure}[t!]
\centering{
\phantomsubcaption\label{fig:multvsage}
\phantomsubcaption\label{fig:accuracyvsage}
}
\includegraphics{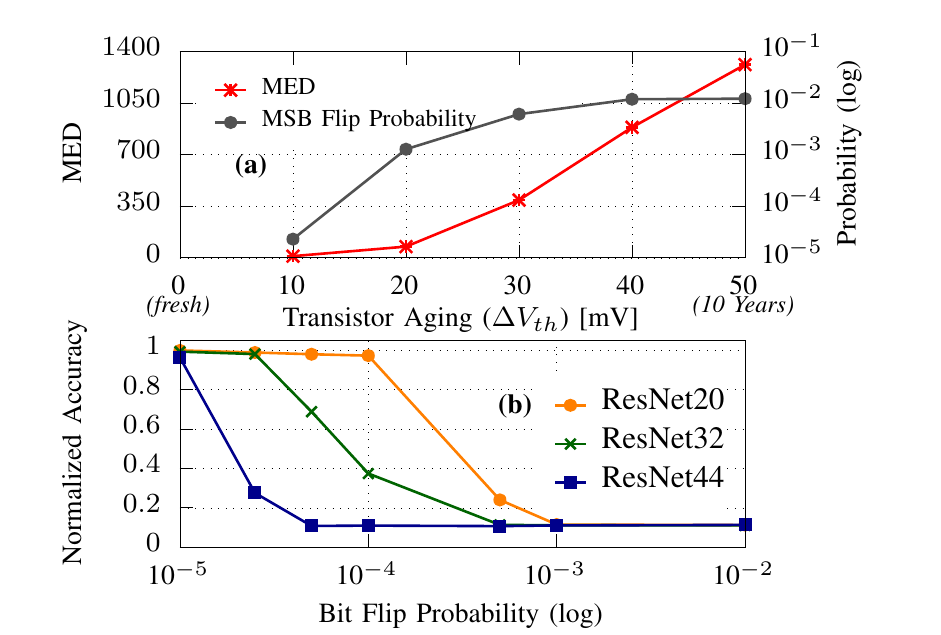}
\caption{a) The error characteristics of an 8-bit multiplier under aging.
b) Accuracy of three ResNets when injecting errors (random bit flips in the 2 MSBs with a given probability) at the multiplications of the convolutional layers.
}
\label{fig:vsage}
\end{figure}

\section{Aging-Induced Timing Errors}
Aging-induced timing degradation increases the circuit's delay over time and thus, timing errors occur when the circuit's paths fail to fulfill the required timing constraints~\cite{amrouch_approximate_aging2} (Eq.~\ref{eq:guardband}).
Induced timing errors result in unacceptable accuracy loss even after just 1 year, as shown in~\cite{amrouch_approximate_aging}.
In arithmetic circuits, errors mainly occur in the most significant bits (MSBs)~\cite{amrouch_approximate_aging}.
%are you sure that this citation claims something similar?
To explore that, we demonstrate in Fig.~\ref{fig:multvsage}, the aging-induced timing errors of a 8-bit multiplier circuit (i.e., the vital component of any NPU).
The circuit is obtained from the commercial Synopsys DesignWare library.
Fig.~\ref{fig:multvsage} presents i) the Mean Error Distance (MED) for different aging levels (represented as different values in threshold voltage increase ($\Delta V_{th}$)) and ii) the resulting probability of a bit flip in one of the two MSBs.
One million random generated inputs are used.
MED is defined as the average absolute distance between the output of the accurate and of the aged circuit.
$\Delta V_{th}$=$50$mV represents the end of lifetime, i.e., 10 years~\cite{ThirunavukkarasuTED2019}.
\textit{No timing guardbands are used in this investigation.
Hence, the multiplier is clocked with the maximum frequency obtained from operation at the critical path delay of the fresh multiplier}.
As shown in Fig.~\ref{fig:multvsage}, more errors are produced over time (higher MED), and the probability of a bit-flip at the MSBs increases significantly.

Recent research demonstrated that the deeper NNs become, the more susceptible to errors generated in the multiplier units their accuracy becomes~\cite{TasoulasTCASI2020}.
In Fig.~\ref{fig:accuracyvsage}, we present an estimation of how aging-induced timing errors impact the accuracy of different NNs.
Note, post-synthesis timing simulation to capture the accuracy loss due to aging is infeasible.
DNNs feature millions of multiplications~\cite{mrazek2019alwann,TasoulasTCASI2020} and simulating only one inference requires several hours~\cite{TasoulasTCASI2020}.
Hence, we run inference at software level and we inject errors in the performed multiplications.
Error injection is implemented by randomly flipping one of the two MSBs with a given probability, where each experiment is performed ten times to capture the average accuracy.
Three ResNets are considered and the probability of the bit flip ranges from $10^{-5}$ to $10^{-2}$.
As shown in Fig.~\ref{fig:accuracyvsage}, as the probability of a bit flip increases, the accuracy drops significantly and becomes unacceptable after a very small probability level of $5\times10^{-4}$.
In addition, for the deeper NNs (i.e., as the size of ResNet increases) the accuracy drops faster.
It is noteworthy that the bit flip probability (\texttildelow $10^{-3}$) of the 8-bit multiplier with only a $20mV$ voltage threshold increase due to aging, results in an unacceptable accuracy drop for all the examined NNs.
\blue{In Fig.~\ref{fig:vsage}, for the sake of exemplification, we examine only the bit-flip probability at the MSBs.}
As shown in Figs.~\ref{fig:vsage}, aging-induced timing errors are critical for the NPU's accuracy, leading to poor quality even after a small degradation.

\section{MAC Delay Analysis Under Input Compression}\label{sec:delay}
The core component of NPUs is the MAC unit.
NPUs perform millions of MAC operations per inference and the overall performance of the NPU is defined by the speed of single MAC unit.
In our work, we consider a microarchitecture based on Google Edge TPU~\cite{cass2019taking} that comprises a $64\times 64$ systolic MAC array.
Hence, we design a MAC unit that consists of an $8$-bit unsigned multiplier and a $22$-bit unsigned adder to prevent accumulation overflow.
Then, we synthesize the RTL description of the MAC using Synopsys Design Compiler targeting maximum performance using the $14$nm FinFET technology which is calibrated with measurement data from Intel.
The optimized arithmetic modules of the DesignWare library are used to describe the MAC unit, as done in commercial chips.

Not all timing paths in a circuit feature the same delay and therefore, the paths that will be \textit{activated} are input-dependent.
For arithmetic circuits (such as a MAC unit), lower bit-width numbers lead to shorter carry propagations and thus, faster operation~\cite{amrouch_approximate_aging,amrouch_approximate_aging2}.
\textit{Exploiting the latter, we compress the inputs of the MAC to achieve faster operation}.
The accurate MAC, in practice, performs the operation: $A \times B + C$.
The width of $A$ and $B$ is $8$ bits while the width of $C$ is $22$ bits.
The compressed inputs $A^\prime$, $B^\prime$, and $C^\prime$ feature a bit-width of $8-\alpha$, $8-\beta$, and $22-(\alpha+\beta)$, respectively.
Since the compressed inputs feature smaller bit-width, we apply zero-padding to the remaining bits.
Two options are available for zero-padding: i) MSB-padding, i.e., fill with zeros the MSB positions, and ii) LSB-padding, i.e., fill with zeros the LSB positions.
In the latter case, the result $A^\prime \times B^\prime + C^\prime$ is shifted left $\alpha+\beta$ places. 

\begin{figure}[t!]
\centering
\includegraphics{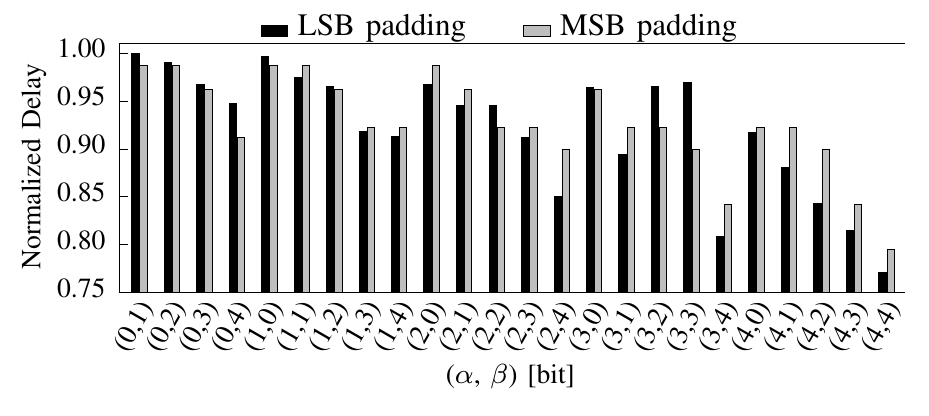}
\caption{Delay gain of the $8$-bit MAC when applying ($\alpha$, $\beta$) input compression.
Both MSB and LSB paddings are evaluated.
}
\label{fig:delayvsize}
\end{figure}

In Fig.~\ref{fig:delayvsize}, we evaluate the delay of our MAC unit when compressing its inputs, i.e., performing $A^\prime \times B^\prime + C^\prime$ instead of $A \times B + C$.
Various compression values ($\alpha$, $\beta$) and both padding options are examined. 
As shown in Fig.~\ref{fig:delayvsize}, around $23\%$ delay gain can be achieved for up to  ($4$, $4$) compression.
In addition, Fig.~\ref{fig:delayvsize} shows that some compression values are benefited by MSB padding while others by LSB padding.
Therefore, both padding options should be considered.
Fig.~\ref{fig:delayvsize} demonstrates that by just compressing the MAC inputs, we can achieve significant delay gain without any circuit modifications. 

In order to compress the MAC inputs, while merely impacting the inference accuracy of the NN, we employ multiple low bit-width quantization techniques\cite{krishnamoorthi2018quantizing,jacob2018quantization,banner2019post,nahshan2019loss}. Particularly, we quantize the activations and the weights to $8-\alpha$ and $8-\beta$ bits respectively, and we perform the appropriate padding afterward (Section~\ref{sec:quantization}). In that way, our approach enables i) accurate operation when no aging effects appear (i.e., no compression), and ii) gradually increase the compression ($\alpha$ and $\beta$ values) over time to increase the delay gain as the NPU ages. 

\section{Input Compression Through Quantization}\label{sec:quantization}
Typically, NNs are trained using a 32-bit floating point (FP32) number representation.
Post-training quantization aims at reducing the number of bits used to represent weights and activations of NNs (e.g., 8-bit integers, INT8).
Consequently, the model size is reduced while the accuracy is maintained very close to the accuracy of the FP32 model~\cite{krishnamoorthi2018quantizing}.

In this work, we utilize post-training quantization to implement input compression and leverage the respective delay gain proposed in Section~\ref{sec:delay} to address NPU aging. 
To represent the compressed inputs, $A^\prime$, $B^\prime$, and $C^\prime$, we target quantization with different bits for weights and activations.
Therefore, we created a library of multiple low-bit width post-training quantization methods based on recently published approaches.
We integrate multiple methods, since some of them are optimized for specific NNs, or they are optimized for very low precision.
Some other methods require off-line statistics and in some cases the actual quantization is time-consuming due to the multiple optimizations.
Particularly, we have included uniform symmetric~\cite{krishnamoorthi2018quantizing} and asymmetric min/max quantization~\cite{jacob2018quantization}, as well as more sophisticated methods and libraries such as ACIQ~\cite{banner2019post} (w/ and w/o bias correction), and LAPQ~\cite{nahshan2019loss}, which support per-channel bit optimizations, stats-based analysis, and bias correction.
\textit{All these methods do not require NN retraining and allow the utilization of different precision for weights and activations}.
\blue{More details about the employed quantization technique in~\cite{krishnamoorthi2018quantizing,jacob2018quantization,banner2019post,nahshan2019loss}.}
As an example to demonstrate that different methods have different effects on NNs, for ResNet50, LAPQ is the best method for W8A4 (INT8 for weights INT4 for activations) where accuracy drops $1.7\%$ on the ImageNet dataset. For W4A4, ACIQ is the best method with a drop of only $4.2\%$ (corresponding drop of LAPQ is $11.3\%$), verifying its design principle that targets low bit-width quantization. Contrary, VGG13 on ImageNet, LAPQ is the best quantization method both for W8A4 and W4A4 with accuracy drop of only $1\%$ and $4.2\%$ respectively. 

For 8-bit quantization, the activations and weights are quantized to the [$0$, $2^8$) segment, while the biases to [$0$, $2^{16}$).
Considering ($\alpha$, $\beta$) compression, the activations are quantized to [$0$, $2^{8-\alpha}$), the weights to [$0$, $2^{8-\beta}$), and the biases to [$0$, $2^{16-\alpha-\beta)}$).
When considering LSB padding, the convolution inputs are shifted left and thus, convolution operation equals:

{
\begin{equation}
\begin{split}
F_{shifted}=Bias\times 2^{(\alpha+\beta)}+\sum_{\forall j}{\big((A_j\times2^\alpha) \times(W_j\times2^\beta)\big)}\\
=\Big( Bias+\sum_{\forall j}{\big(A_j \times W_j \big)\Big) }\times 2^{(\alpha+\beta)} = F \times 2^{(\alpha+\beta)}
\end{split}
\end{equation}
}

\noindent Hence, the output needs to be shifted right $\alpha+\beta$ positions.
However, the latter does not require additional hardware since it can be easily performed at the software side when storing/reading the convolution result to/from the memory.
On the other hand, when MSB padding is used, no shift is required.

\section{Implementation of Aging-Aware Quantization} \label{sec:our_framework}
\begin{figure}[t!]
\centering
\resizebox{0.37\columnwidth}{!}{\includegraphics{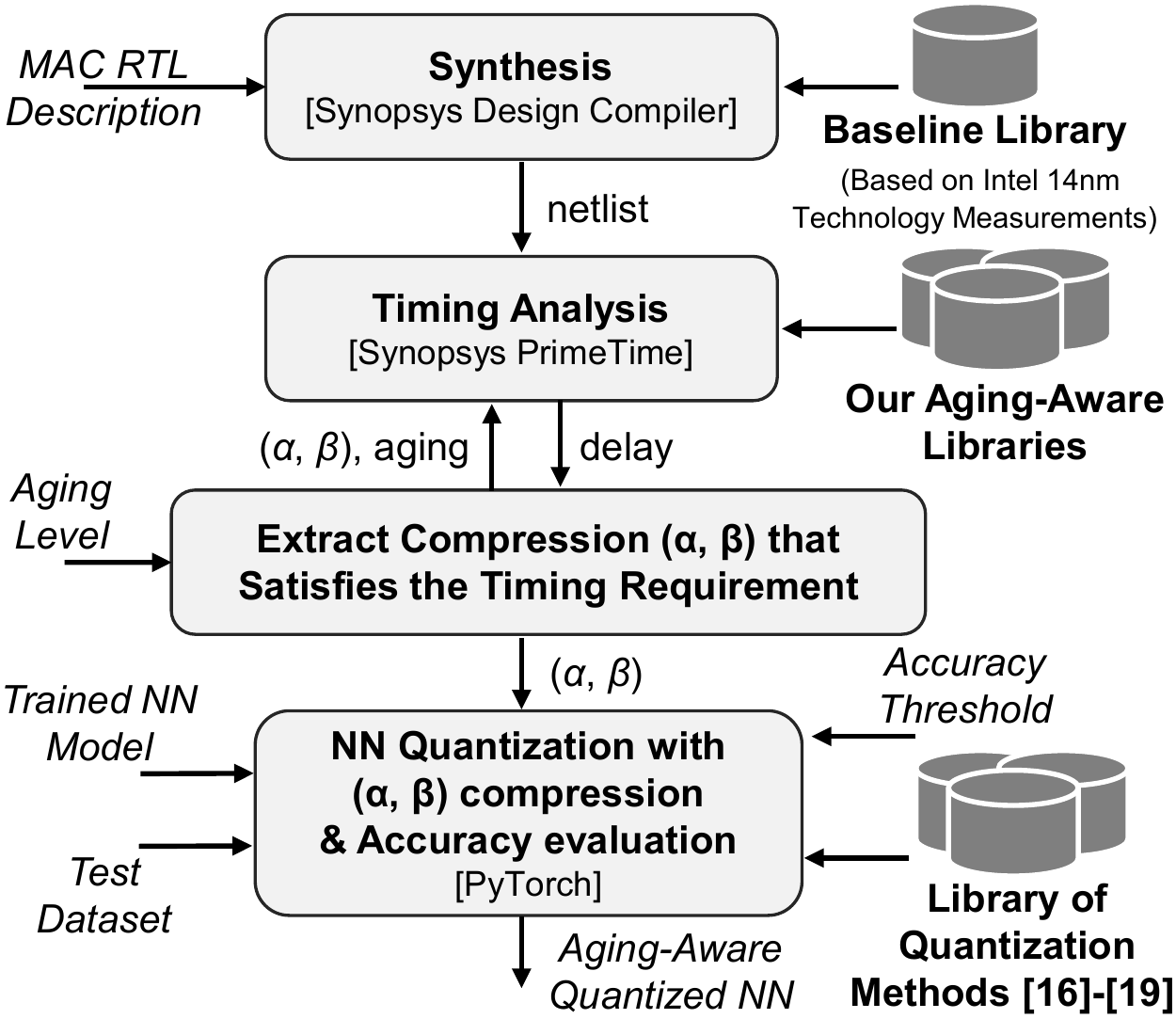}}
\caption{Our device-to-system implementation that enables aging-aware quantization to suppress aging effects in NPUs.}
\label{fig:framework}
\end{figure}

Here, we describe our implementation (summarized in Fig.~\ref{fig:framework}) to perform the proposed aging-aware quantization.
Our implementation starts from the device level all the way up to the system level where the NN inference accuracy is impacted.

\subsection{Aging Modeling}\label{subsec:agemodel}
\begin{enumerate}[wide, labelindent=0pt,label=\textbf{(\arabic*)}, ref=(\arabic*)]
\item \textbf{Aging Model:}
In this work, we employ the state-of-the-art physics-based aging model~\cite{PariharTED2018}.
This model has been validated against semiconductor measurements for various technologies and transistor structures. 
It is able to precisely capture aging-induced degradation ($\Delta V_{th}$) for the $14$nm technology.

\item \textbf{Aging-Aware Cell Libraries:}\label{item:libs}
To enable aging support in commercial digital design tool flows, we generate aging-aware cell libraries.
First, we calibrate the (BSIM-CMG) model to match Intel's $14$nm FinFET technology measurements provided in~\cite{intel_data}. Details on calibration and validation of BSIM-CMG are available in~\cite{mishra2018simulation}.
Then, we employ the state-of-the-art physics-based aging model to estimate the corresponding $\Delta V_{th}$ over time. 
In this work, we consider aging from 0 to 10 years for the typical projected lifetime.
Aging gradually increases $\Delta V_{th}$ over time.
$\Delta V_{th}$ for a fresh chip is equal to $0$, while after 10 years it reaches $50$mV, as demonstrated from measurements for the FinFET technology~\cite{ThirunavukkarasuTED2019,PariharTED2018}.
Aging mechanism is affected by the operating conditions (e.g., utilization and temperature).
For instance, $\Delta V_{th}=20$mV may correspond to 1-2 years.
Hence, in our analysis, we consider $\Delta V_{th}$ as an unbiased measure of the aging level and we investigate aging effects at different $\Delta V_{th}$ levels from $0$mV (fresh) to $50$mV (10 years) with a step of $10$mV.  
Afterward, for each $\Delta V_{th}$ step, we create an aging-aware cell library by characterizing all standard cells on the respective $\Delta V_{th}$ based on the open-source FinFET standard cells from Silvaco, using Synopsys SiliconSmart.
This is done using the commercial SPICE simulation that measures the delay and power of every standard cell under the influence of $\Delta V_{th}$. 

\item  \textbf{Aging-induced Delay Analysis:}\label{item:delay}
Synopsys Design Compiler and the \texttt{compile\_ultra} command are used to synthesize the circuit's RTL description targeting maximum performance, i.e., zero-slack.
For synthesis, we consider the fresh library, i.e., baseline without aging.
Next, we use Synopsys PrimeTime to perform static timing analysis (STA) on the post-synthesis netlist.
During STA we employ our aging-aware cell libraries to precisely capture the impact of aging on the circuit's delay.
\blue{We consider the worst-case analysis where all transistors exhibit the maximum degradation.}
In addition, we analyze the timing information of both the uncompressed and compressed (i.e., reduced bit width) inputs.
In the latter case, we specify that the respective input bits of the bit positions that are padded with zeros (due to input compression) are constantly set to `0'.
\blue{Hence, we precisely obtain the circuit's delay w.r.t. the aging period and the paths that are activated due input compression.} 
\end{enumerate}

\subsection{Our Proposed Aging-Aware Quantization}
\label{subsec:aging_aware_adaptive_app}
In Section~\ref{sec:delay}, we demonstrated that applying input compression on the MAC unit delivers considerable delay gains that can potentially eliminate the aging-induced timing errors.
As discussed in Section~\ref{sec:quantization}, input compression in NPUs can be applied by employing low bit-width quantization.
Nevertheless, the latter still results in accuracy loss, despite the numerous quantization methods that have been proposed.
In our work, we introduce an adaptive approximation approach by progressively increasing the input compression over time.
Our implementation is illustrated in Fig.~\ref{fig:framework} and described in Algorithm~\ref{alg:framework}.
First, we synthesize the RTL description of our circuit as described above (Section~\ref{subsec:agemodel}~\ref{item:delay}) to obtain the post-synthesis netlist  without aging.
We consider the MAC unit as our driving circuit since the accuracy and delay of the MAC operation will define the accuracy and speed of the NPU~\cite{jouppi2017datacenter,TasoulasTCASI2020}.
Next, we use PrimeTime and our aging-aware libraries (Section~\ref{subsec:agemodel}~\ref{item:libs}) to perform timing analyses to identify all the compression values ($\alpha$, $\beta$), under both MSB and LSB padding, that satisfy the timing constraint of the MAC unit (lines 2-4).
%In the timing analysis we consider both MSB and LSB padding.
Targeting minimum compression, we select the ($\alpha$, $\beta$) that minimizes $\sqrt{\alpha^2+\beta^2}$ (line 5).
In the case of a tie, we select the ($\alpha$, $\beta$) with the highest precision for the activations~\cite{banner2019post} (i.e., smallest $\alpha$).
Finally, we use the obtained compression value ($\alpha$, $\beta$) to  quantize the NN model.
The quantization size equals $8-\alpha$ for the activations, $8-\beta$ for the weights, and $16-\alpha-\beta$ for the biases.
For the quantization procedure, we consider all of the available methods in our library (Section~\ref{sec:quantization}), and we capture the inference accuracy on the test dataset by using the quantized model (lines 6-8).
If a user-defined accuracy loss threshold is satisfied, then the quantized model is the output of our algorithm (line 9).
If a desired accuracy loss threshold is not available, we iterate over all the quantization methods to select the one that delivers the highest accuracy.

\begin{algorithm}[t!]
\caption{Aging-Aware Quantization}
\label{alg:framework}
\begingroup
\begin{tabular}{l l}
\textbf{Input:}& 1. Synthesized Netlist, 2. Aging Level: e.g., $\Delta V_{th}$, \\
               & 3. Trained Model \& Test Dataset, 4. Accuracy Loss Threshold: $e$ \\
\textbf{Output:} & Aging-Aware Quantized Model
\end{tabular}
\endgroup
\begin{algorithmic}[1]
\State List = []
\State \textbf{for all} ($\alpha$, $\beta$) $\in$ [$0$, $8$]$^2$:
\State \hspace{2mm} Run STA with ( corresponding aging library, compression ($\alpha$, $\beta$) )
\State \hspace{2mm} \textbf{if} timing constraint is met\textbf{:} List $\leftarrow$ add ($\alpha$, $\beta$)
\State ($\alpha$, $\beta$) $\leftarrow$ ($\alpha$, $\beta$) in List with min\big($\sqrt{\alpha^2+\beta^2}$\big)
\State \textbf{for all} method \textbf{in} Quantization Library
\State \hspace{2mm} Quantize Model using method and size (8-$a$, 8-$b$)
\State \hspace{2mm} Capture accuracy on test dataset
\State \hspace{2mm} \textbf{if} threshold $e$ is satisfied\textbf{:} \textbf{return} Quantized Model
\end{algorithmic}
\end{algorithm}

The ($\alpha$, $\beta$) values that are extracted during the timing analysis phase ensure that the timing constraint is met.
Hence, no aging-induced timing errors occur and accurate computations are performed on the compressed inputs.
In that way, the inference accuracy is defined \textit{only} by the accuracy delivered by quantization for the respective compression values.
\textit{Therefore, the inference accuracy can be captured purely at software level, without the need to perform time-consuming post-synthesis timing simulations that are infeasible for large datasets}~\cite{TasoulasTCASI2020}.
In addition, the padding selection does not affect the quantization process/accuracy, and only affects how the data will be stored in memory (see Section~\ref{sec:quantization}).
Note that the $\alpha$ and $\beta$ values depend on the NPU microarchitecture (e.g., MAC size) and the aging period.
On the other hand, the selected quantization method depends on both the ($\alpha$, $\beta$) and the NN.
Hence, for a specific aging period, different NNs will feature the same ($\alpha$, $\beta$) while they might utilize a different quantization method.
Targeting to minimize the employed compression, in line 5 of Algorithm~\ref{alg:framework}, we select the  ($\alpha$, $\beta$) that minimizes $\sqrt{\alpha^2+\beta^2}$, i.e., we use the Euclidean distance from ($0$, $0$) as a surrogate model of the applied compression.
To evaluate the efficiency of our model in estimating the applied compression we run the following experiment.
For each quantization method in our library and for each NN examined in Section~\ref{sec:experimental}, we quantize the NN using the respective method and ($\alpha$, $\beta$) compression and then, we capture the accuracy loss w.r.t. the FP32 model.
This procedure is repeated $\forall$($\alpha$, $\beta$) $\in$ $[0,4]^2$.
Next, we rank ($\alpha$, $\beta$) based on i) the computed accuracy loss, and ii) our model.
Finally, we calculate the Pearson Correlation Coefficient between the two rankings obtained.
Over the ten examined NNs and the five quantization methods, the Pearson Coefficient is $0.84$ on average (ranging from $0.71$ to $0.92$).
Hence, our model achieves a very strong correlation in ranking the ($\alpha$, $\beta$) compression values.
As an alternative, we can evaluate lines 6-8 for all the extracted ($\alpha$, $\beta$) values.
However, considering that some quantization methods are time-consuming, this would heavily impact the execution time of our implementation.
On the other hand, by considering only one ($\alpha$, $\beta$) during quantization, only $1$ hour was required, in the worst case.

\begin{table}[!]
\renewcommand{\arraystretch}{1.2}
\caption{Achieved accuracy and selected quantization method for varying NNs at various aging levels (represented by $\Delta V_{th}$).}
\label{tab:accuracy}
\centering
\begin{threeparttable}
\footnotesize
\begin{tabular}[t]{l|c|c|c|c|c}
  \hline
 & \multicolumn{5}{c}{\textbf{Accuracy Loss (\%) / Quantization Method Selected}}\\ \hline
\textbf{Neural Network} & \textbf{$\boldsymbol{10}$mV} & \textbf{$\boldsymbol{20}$mV} & \textbf{$\boldsymbol{30}$mV} & \textbf{$\boldsymbol{40}$mV} & \textbf{$\boldsymbol{50}$mV} \\ \hline
ResNet50 & 0.27 / M5\tnote{*} & 0.36 / M5 & 0.97 / M3\tnote{*} & 1.47 / M4\tnote{*} & 2.37 / M4\\ \hline
ResNet101 & 0.26 / M5 & 0.36 / M5 & 1.28 / M5 & 0.97 / M4 & 1.84 / M4\\ \hline
ResNet152 & 0.28 / M5 & 0.34 / M5 & 1.08 / M5 & 1.12 / M4 & 2.10 / M4\\ \hline
VGG13 & 0.15 / M4 & 0.22 / M3 & 0.39 / M3 & 1.20 / M4 & 2.54 / M4\\ \hline
VGG16 & 0.05 / M5 & 0.14 / M5 & 0.29 / M3 & 0.73 / M4 & 1.09 / M4\\ \hline
VGG19 & 0.20 / M3 & 0.33 / M3 & 0.46 / M3 & 1.09 / M4 & 2.37 / M4\\ \hline
Alexnet & 0.28 / M5 & 0.54 / M5 & 0.99 / M5 & 2.72 / M4 & 4.00 / M4\\ \hline
SqueezeNet 1.1 & 0.55 / M5 & 1.51 / M5 & 3.61 / M4 & 6.03 / M4 & 7.83 / M4\\ \hline
Wide ResNet50 & 0.14 / M5 & 0.24 / M5 & 0.67 / M5 & 1.27 / M4 & 2.49 / M4\\ \hline
Wide ResNet101 & 0.23 / M5 & 0.41 / M5 & 1.33 / M5 & 1.41 / M4 & 2.92 / M4\\ \hline
\end{tabular}
\begin{tablenotes}\footnotesize
\item[*] M3: LAPQ~\cite{nahshan2019loss}, M4: ACIQ~\cite{banner2019post}, M5: ACIQ w/o bias correction~\cite{banner2019post}
\end{tablenotes}
\end{threeparttable}
\end{table}

\begin{table}[!]
\renewcommand{\arraystretch}{1.2}
\caption{The extracted compression values ($\alpha$,$\beta$) and padding for the examined aging levels.}
\label{tab:compression}
\centering
\footnotesize
\begin{tabular}[t]{l|c|c|c|c|c}
  \hline
 \textbf{Aging [$\boldsymbol{\Delta V_{th}}$]} & \textbf{$\boldsymbol{10}$mV} & \textbf{$\boldsymbol{20}$mV} & \textbf{$\boldsymbol{30}$mV} & \textbf{$\boldsymbol{40}$mV} & \textbf{$\boldsymbol{50}$mV} \\ \hline
\textbf{($\boldsymbol{\alpha}$, $\boldsymbol{\beta}$) / Padding} & (2,0)/LSB & (2,2)/MSB & (3,1)/LSB & (2,4)/LSB & (3,4)/LSB) \\ \hline
\end{tabular}
\end{table}

\section{Experimental Results and Evaluation}\label{sec:experimental}
In order to evaluate the effectiveness of our technique in eliminating the aging-induced timing errors in NPUs, we examine the delay gain delivered by our technique as well as the respective accuracy loss that has to be traded due to the applied input compression.
For our evaluation, we consider the Edge TPU microarchitecture~\cite{cass2019taking} and we use the MAC unit described in Section~\ref{sec:delay} (i.e., $8$-bit multiplier/$22$-bit adder) as our driving circuit.
In addition, we consider ten NNs (listed in Table~\ref{tab:accuracy}) with varying characteristics.
For aging analysis, we use the libraries and workflow described in Section~\ref{subsec:agemodel}.
All the NNs are trained on the ImageNet dataset~\cite{deng2009imagenet} and their implementation is based on official PyTorch repositories (Torchvision)~\cite{m:pytorch}.
Hereafter, when referring to the baseline design, we refer to the MAC unit when using $8$-bit quantization for activations and weights~\cite{jouppi2017datacenter} (i.e., no compression $\alpha$=$\beta$=$0$).
Moreover, the accuracy loss is calculated with respect to the accuracy achieved with FP32 inference.
Therefore, even the baseline with no-aging will exhibit a small (negligible) accuracy loss.
In our algorithm, we do not set an accuracy loss threshold but instead, we iterate over all the quantization methods to select the one that delivers the highest accuracy.
For aging, we examine 10 years as the typical projected lifetime.
Finally, for our analysis, we also evaluated precision scaling by applying LSB masking on the $8$-bit quantized NNs~\cite{amrouch_approximate_aging,amrouch_approximate_aging2}.
However, without retraining precision scaling delivered unacceptable accuracy loss for all the examined NNs and aging levels.
Considering, that DNN retrain is very time consuming~\cite{TasoulasTCASI2020}, precision scaling~\cite{amrouch_approximate_aging,amrouch_approximate_aging2} is not included in our discussion.

\begin{figure}[t!]
\centering{
\phantomsubcaption\label{fig:evaldelay}
\phantomsubcaption\label{fig:evalacc}
}
\centering
\includegraphics{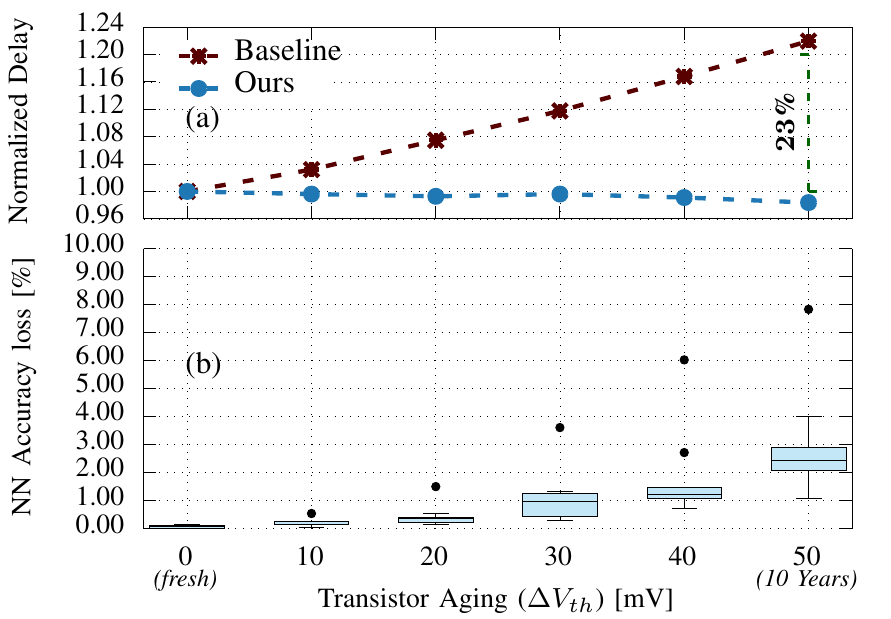}
\caption{a) The normalized delay, from the beginning (i.e., fresh) to the end of lifetime (10 years), of the baseline and our approach w.r.t the delay of the fresh baseline. 
b) Graceful accuracy degradation, over time, delivered by our aging-aware quantization, presented by box plots w.r.t. the examined NNs.
}
\label{fig:evaluation1}
\end{figure}

Table~\ref{tab:compression} reports the compression values ($\alpha$,$\beta$) and padding as extracted by our algorithm for the examined aging periods (i.e., aging is represented by an increasing $\Delta V_{th}$ over time).
Over the aging periods, Fig.~\ref{fig:evaldelay} depicts the delay of the MAC when employing our technique (i.e., applying inputs compression as in Table~\ref{tab:compression}) in comparison with the delay degradation of the baseline.
In Fig.~\ref{fig:evaldelay}, the delay is normalized with respect to the delay of the fresh baseline (i.e., without aging) without timing guardbands.
As shown, the delay of the baseline increases over time (normalized delay is higher than $1$), reaching a performance loss of $23\%$ under $10$ years aging.
This is an expected behavior due to the aging phenomena where transistors become slower.
Hence, without timing guardbands, aging-induced timing errors occur which will significantly deteriorate the NPU accuracy (see Fig.~\ref{fig:accuracyvsage}).
Contrarily, this is not the case when applying our technique.
As Fig.~\ref{fig:evaldelay} demonstrates, our approach does not follow the baseline degradation style.
\blue{\textit{This is because our technique adaptively compresses the inputs over time to gracefully compensate the delay increase due to aging.}}
The normalized delay is always less than or equal to $1$, making our aging-aware quantization technique resilient against aging-induced delay degradation.
Therefore, considering that the achieved delay can be always similar, if not even better, to the delay without aging, our technique effectively suppresses the aging effects and no timing errors will occur. 
Importantly, our technique shows almost fixed delay from day-zero until the end of the projected lifetime, and hence no guardband is required.
Thus, for 10 years lifetime, by removing the timing guardband, a $23\%$ delay gain is achieved compared to baseline.

Nevertheless, our approach results in accuracy degradation since lower range is used for the weights and activations.
For the examined NNs, Fig.~\ref{fig:evalacc} depicts the accuracy degradation over aging when applying our aging-aware quantization (see Table~\ref{tab:compression}).
\blue{The ($\alpha$,$\beta$) and padding in Table~\ref{tab:compression} refer to worst-case delay analysis and hence their selection depends solely on the aging period.
Thus, the same compression is used for all the NNs.}
Note that, since our technique eliminates the timing errors, the accuracy of the aged NPU matches the accuracy achieved by the respective exact quantized model (i.e., ($\alpha$,$\beta$) compression and no-aging).
In Fig.~\ref{fig:evalacc}, the accuracy loss is presented by box plots over the examined NNs and aging periods.
As shown, our technique delivers graceful accuracy degradation over time.
For example, the average accuracy loss is $0.24$\%, $0.45$\%, $1.11$\%, $1.80$\%, and $2.96$\% for aging ($\Delta V_{th}$) $10$mV, $20$mV, $30$mV, $40$mV, and $50$mV, respectively, where $50$mV is equal to 10 years aging.
In addition, as shown in Fig.~\ref{fig:evalacc}, for all the examined periods, the accuracy loss is well concentrated around the median demonstrating that our technique is slightly affected by the NN model.
The highest reported accuracy loss is $7.83$\% for 10 years aging for the SqueezeNet network.
SqueezeNet features always the highest accuracy loss for all aging periods.
SqueezeNet is by design a very compressed network and thus, it is affected considerably by low bit-width quantization.
\blue{The above analysis examines timing guardband elimination.
However, with ($3$,$1$) compression and only $9$\% guardband the accuracy loss becomes $1.11$\%, on average, for 10 years aging.}
The full accuracy results are summarized in Table~\ref{tab:accuracy}.
In addition, in Table~\ref{tab:accuracy}, we report the quantization method selected by our algorithm for each case.
\blue{Table~\ref{tab:accuracy} reports the accuracy as obtained from PyTorch for the respective compression and quantization method.}
As shown in Table~\ref{tab:accuracy}, the quantization method LAPQ~\cite{nahshan2019loss} is selected in the $14$\% of the cases, while ACIQ~\cite{banner2019post} and ACIQ w/o bias~\cite{banner2019post} are selected in the $44$\% and $42$\% of the cases, respectively.
The methods~\cite{krishnamoorthi2018quantizing,jacob2018quantization} were not selected in any aging level since the required compression values (Table~\ref{tab:compression}) were very high and out of the effective range of~\cite{krishnamoorthi2018quantizing,jacob2018quantization}.
Table~\ref{tab:accuracy} highlights the importance of considering a quantization methods library since the best quantization method varies with respect to the required compression and the NN model itself.

\begin{figure}[t!]
\centering
\includegraphics{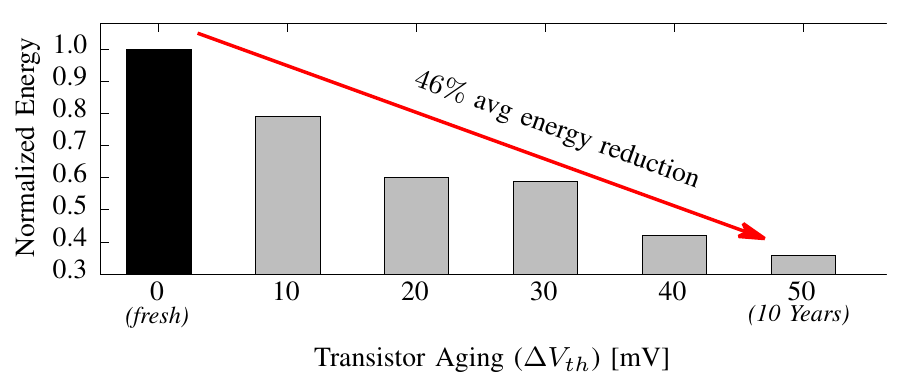}
\caption{Normalized energy consumption of our technique over the baseline for varying aging levels. Aging timing guardband is used for the baseline to prevent timing errors.
}
\label{fig:energy_sav}
\end{figure}

Finally, we evaluate the energy efficiency of our technique.
Over time, our technique  adaptively compresses the MAC inputs to eliminate the aging effects.
However, input compression leads also to reduced switching activity and thus, lower power consumption.
Fig.~\ref{fig:energy_sav} depicts the normalized energy of our technique w.r.t the energy consumption of the baseline for varying aging levels.
In Fig.~\ref{fig:energy_sav}, the MAC unit using our approach is operated at the maximum frequency of the fresh MAC, while for the baseline a 23\% timing guardband (Fig.~\ref{fig:evaldelay}) is used to prevent timing errors due to aging.
As shown in Fig.~\ref{fig:energy_sav}, for no aging, our technique does not induce any overhead, while for $10$mV-$50$mV aging, it delivers 46\% energy reduction on average (ranging from 21\% up to 67\%).

\section{Conclusion}
In this paper, we employ quantization, for the first time, to suppress aging in NPUs.
Our technique applies adaptive aging-aware quantization over time and achieves a progressive, though small, accuracy degradation, while in the meantime it eliminates the deleterious aging-induced timing errors.
Our technique enables removing the aging timing guardbands and boosts the NPU performance, while in the meantime it significantly decreases the energy consumption.
Moreover, our approach modifies only the inputs of the NPU and it does not require any modifications to the underlying microarchitecture.

\section*{Acknowledgement}
Authors would like to thank 
Yogesh Chauhan and Souvik Mahapatra and their teams for the valuable help in compact model calibration. 
This work is supported in part by the German Research Foundation (DFG) through the project ``ACCROSS: Approximate Computing aCROss the System Stack''.

\end{document}